\documentclass[fleqn,usenatbib]{mnras}

\usepackage{newtxtext,newtxmath}
\usepackage[T1]{fontenc}

\DeclareRobustCommand{\VAN}[3]{#2}
\let\VANthebibliography\thebibliography
\def\thebibliography{\DeclareRobustCommand{\VAN}[3]{##3}\VANthebibliography}

\usepackage[usenames]{xcolor}
\usepackage{graphicx}	% Including figure files
\usepackage{amsmath}	% Advanced maths commands
\usepackage{comment}
\usepackage{booktabs,gensymb}
\usepackage{multirow}
\usepackage{leftidx}
\usepackage{natbib}
\usepackage{lastpage}

\newcommand{\rev}[1]{#1}
\newcommand{\revv}[1]{#1}
\newcommand{\revvv}[1]{#1}

\newcommand{\Msun}{\,M_{\odot}}

\newcommand{\epsff}{\epsilon_{\mathrm{ff}}}

\newcommand{\lambdam}{\lambda_{\mathrm{m}}}

\title[Disruption of star clusters]{Tidal disruption of star clusters in galaxy formation simulations}

\author[Meng et al.]{
Xi Meng \thanks{E-mail: xim@umich.edu}%
\href{https://orcid.org/0000-0002-8276-4164}{\includegraphics[scale=0.6]{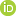}} and 
Oleg Y. Gnedin \href{https://orcid.org/0000-0001-9852-9954}{\includegraphics[scale=0.6]{orcid.png}}
\\
Department of Astronomy, University of Michigan, Ann Arbor, MI 48109, USA
}

\date{Accepted XXX. Received YYY; in original form ZZZ}

\pubyear{2021}

\begin{document}
\label{firstpage}
\pagerange{\pageref{firstpage}--\pageref{lastpage}}
\maketitle

\begin{abstract}
We investigate the evolution of the tidal field experienced by massive star clusters using cosmological simulations of Milky Way-sized galaxies.
Clusters in our simulations experience the strongest tidal force in the first few hundred Myr after formation, when the maximum eigenvalue of the tidal tensor reaches several times $10^4$~Gyr$^{-2}$. 
After about 1~Gyr the tidal field plateaus at a lower value, with the median $\lambdam \sim 3\times10^3$~Gyr$^{-2}$. 
The fraction of time clusters spend in high tidal strength ($\lambdam > 3\times10^4$~Gyr$^{-2}$) regions also decreases with their age from $\sim$20\% immediately after formation to less than 1\% after 1~Gyr. 
At early ages both the {\it in situ} and {\it ex situ} clusters experience similar tidal fields, while at older ages the {\it in situ} clusters in general experience stronger tidal field due to their lower orbits in host galaxy. 
This difference is reflected in the survival of clusters: we looked into cluster disruption calculated in simulation runtime and found that {\it ex situ} star clusters of the same initial mass typically end up with higher bound fraction at the last available simulation snapshot than the {\it in situ} ones.
\end{abstract}


\section{Introduction}

 All massive galaxies ($M_* \gtrsim 10^9\Msun$) in the local universe host globular cluster (GC) populations \citep{Brodie:2006aa}. 
 The overlapping metallicity, density, and mass distributions between GCs and young star clusters suggest that they share similar formation mechanisms, except that they differ in their cosmological history \citep{Krumholz:2019aa}. 
 Observations show that the combined mass of the GC system scales almost linearly with the host galaxy halo mass \citep{Spitler:2009aa,Hudson:2014aa,Harris:2017aa,Forbes:2018aa}. 
 This suggests that GC systems are tightly linked to the assembly history of their host halos.
 Therefore, studying the formation and evolution of GC systems can help us understand galaxy evolution better.

 From a newly formed population of star clusters to old GCs there is a significant decrease in the fraction of low-mass clusters. 
 To understand how the power-law initial mass function of young clusters transforms into the lognormal mass function of GCs at present \citep[e.g.][]{Kruijssen:2009aa,Muratov:2010aa,Li:2014aa}, we must understand the dynamical evolution of star clusters. 
 There are internal processes, including stellar evolution and relaxation, as well as external processes, namely tidal disruption, that affect cluster properties. 
 The mass loss due to stellar evolution for single stellar populations \citep{Leitherer:1999aa} and the two-body relaxation for isolated clusters \citep{spitzer87} are well studied.
 However, since tidal disruption involves changing gravitational potential over the cosmic time, complicated orbits of clusters, encounters with massive objects, interactions with the dense disc etc., modelling the disruption effects on star clusters is not straightforward.

 Analytical and semi-analytical models of tidal disruption usually assume a fixed average tidal field without considering the position information  \citep[e.g.,][]{Boylan-Kolchin:2017aa,choksi_etal18,Choksi:2019aa,El-Badry:2019aa} or an analytical tidal field along idealised elliptical orbits in the host galaxy  \citep[e.g.,][]{prieto_gnedin08,Webb:2014aa,Rossi:2016aa}.
 Either assumption does not capture the cosmic evolution or the granularity of the gravitational potential.
 There have also been several studies that used numerical simulations of galaxy formation and galaxy mergers to obtain more realistic tidal fields for modelling of GCs disruption. 
 These studies select a subset of stellar particles to represent GCs orbits.
 For example, \citet{Renaud:2017aa} selected particles older than 10~Gyr in a cosmological zoom-in simulation of a MW-like galaxy and found that the tidal field grows stronger with cosmic time and that the \textit{in situ} clusters experience significantly stronger tides than the accreted clusters. 
 \citet{Halbesma:2020aa} applied the same selection to the Auriga simulation suite but found that it cannot match the observed Milky Way (MW) and M31 GC populations.
 The E-MOSAICS simulations explored cluster formation times, mass and metallicity distributions, orbits and assembly of the GC system \citep[e.g.][]{Reina-Campos:2019aa,Reina-Campos:2020aa,Bastian:2020aa,Keller:2020aa,Pfeffer:2020aa,Trujillo-Gomez:2021aa}. 
 These relations can, in turn, be used to infer the merger history of the MW and possibly other galaxies \citep{Kruijssen:2019aa,Kruijssen:2020aa}. 
 There are also merger and interacting galaxy simulations that explored the effects of merger environment on the formation and disruption of star clusters \citep[e.g.][]{Kruijssen:2012aa,Kim:2018aa}. 
 These efforts have greatly contributed to our understanding of GC evolution.

 The previous work can still be improved with more realistic modeling of the formation of massive star clusters. The history of the tidal field experienced by real clusters may be significantly different from that of all old stellar particles.
 In this work, we present a study of the tidal disruption processes in a suite of cosmological simulations that directly model the formation of star clusters.
 We track the position and tidal field of massive star clusters throughout cosmic time from their birth until redshift $z\approx1.5$. 
 We introduce our simulations and tidal field calculations in Section~\ref{sec:sim}.
 In Section~\ref{sec:result} we describe how the tidal field as well as the location of massive clusters evolve with time, and how that affects the bound fraction of the star clusters. 
 In Section~\ref{sec:discuss} we discuss caveats in our analysis and compare with other studies. 
 We also discuss possible implications of our results for building more realistic analytical models, as well as recovering the assembly history of the MW. 
 We present our conclusions in Section~\ref{sec:conclude}. 

\section{Simulations} \label{sec:sim} 

 We use a suite of cosmological simulations run with the Adaptive Refinement Tree (ART) code  \citep{Kravtsov:1997aa,Kravtsov:1999aa,Kravtsov:2003aa,Rudd:2008aa} and described in \citet{Li:2018aa} and \citet{Meng:2019aa}. 
 All runs start with the same initial conditions in a periodic box of 4 comoving Mpc, producing a main halo with total mass $M_{200}\sim10^{12}\Msun$ at $z=0$, similar to that of the Milky Way. 
 The ART code uses adaptive mesh refinement to achieve better spatial resolution in denser regions. 
 The lowest resolution level is set by the root grid, which is $128^3$ cells. 
 This sets the dark matter (DM) particle mass $m_{\rm DM}=1.05\times10^6\Msun$. 
 The finest refinement level is set to be kept between 3 and 6 physical pc. 
\rev{The ART code uses adaptive mesh refinement, and each particle (stellar particle and DM particle) contributes to the gravitational potential and feels the potential via the cell that it is in. The gravitational softening length is essentially the cell size, which is adaptive according to a combination of the Lagrangian and the Jeans refinement criteria. For details of the refinement criteria we refer the readers to Section~2 of \citet{Li:2017ab}. The maximum refinement level of DM particles is 4 levels above the maximum refinement level of gas cells. }

 The simulations include three-dimensional radiative transfer using the Optically Thin Variable Eddington Tensor approximation \citep{Gnedin:2001aa} of ionizing and ultraviolet radiation from stars \citep{Gnedin:2014aa} and the extragalactic UV background \citep{Haardt:2001aa},  non-equilibrium chemical network that deals with ionization stars of hydrogen and helium, and phenomenological molecular hydrogen formation and destruction \citep{Gnedin:2011aa}. 
 The simulations incorporate a subgrid-scale model for unresolved gas turbulence \citep{Schmidt:2014aa,Semenov:2016aa}.
 
 A unique advantage of these simulations for our study is direct modeling of bound star clusters, rather than generic stellar particles. 
 Star formation is implemented with the continuous cluster formation (CCF) algorithm \citep{Li:2017ab}, where each stellar particle represents a star cluster that forms at a local density peak and grows its mass via accretion of gas until its own feedback terminates its growth. 
 The feedback recipe includes early radiative and stellar wind feedback, as well as a supernova (SN) remnant feedback model \citep{Martizzi:2015aa,Semenov:2016aa}. 
 The momentum feedback of the SN remnant model is boosted by a factor $f_{\rm boost}=5$ to compensate for numerical underestimation due to limited resolution and to match the star formation history expected from abundance matching. 
 The simulations have been run with different values of local star formation efficiency (SFE) per free-fall time, $\epsff$. 
 For a full description of the star formation and feedback recipe, see \citet{Li:2017ab,Li:2018aa}. 
 
 For this analysis we use five runs with different local efficiency: SFE10, SFE50, SFE100, SFE200, and SFE200w. 
 The number after `SFE' in their names is $\epsff$ in percent. 
 There is also run SFE200w with more frequent output of the detailed tidal information for all massive stellar particles ("tidal writeout"). 
 It is a rerun of the SFE200 run, but because of the rather explosive nature of star formation with high SFE, they are not exactly the same \citep{li_gnedin19}.

 We restrict our analysis to massive clusters because their disruption timescale is longer under the same tidal field, giving them greater probability to survive to the present and become GCs.
 We take the threshold of the initial mass of the massive stellar particles to be $M_i > 3\times10^5\Msun$ in all runs except SFE10, where we take $M_i > 2\times10^5\Msun$ because that run has few massive stellar particles due to its low SFE.
 This selection resulted in 40, 73, 46, 121 and 133 star clusters chosen from the SFE10, SFE50, SFE100, SFE200 and SFE200w runs, respectively. 
 
 New massive stellar particles are identified in the halo of the main galaxy in each snapshot and then traced throughout the simulation.
 The center of the galaxy is defined as the location of maximum stellar density, found iteratively using smoothing kernels of decreasing size as in \citet{Brown:2018aa}. 
 The galaxy plane is defined using the shape tensor and its principle axes for neutral gas within 10~kpc from the galaxy center. 
 For full description of the determination of the galaxy plane see \citet{Meng:2019aa}. 
 The cylindrical radius from the galaxy center ($R$) and the height above the disc plane ($z$) are calculated for all massive stellar particles.

 %The ambient matter density ($\rho$) is calculated on the scale of 100~pc as a sum of the gas density in the cell containing the stellar particle, and the ambient density of DM particles and stellar particles.
 %The latter is calculated assuming DM and stellar particles are constant density spheres of radius 100~pc.
\rev{The ambient matter density $\rho$ is calculated on the scale of 100~pc as a sum of the gas density in the cell containing the stellar particle, and the average density of DM particles and stellar particles: $\rho = \rho_{\rm gas} + \rho_{\rm DM} + \rho_*$. The latter two are calculated by representing DM and stellar particles are constant density spheres of radius 100~pc, and summing them within a 100~pc radius sphere around the stellar particle.}

 The tidal tensor around massive clusters is calculated in post-processing of simulation snapshots as the second derivative of the potential: 
 \begin{equation}
  T_{ij}({\textbf x}_0,t) \equiv -\left. \frac{\partial^2\Phi ({\textbf x},t)}{\partial x_i \partial x_j} \right|_{{\textbf x}={\textbf x}_0}
 \end{equation}
 where $i$ and $j$ are two orthogonal directions in the Cartesian coordinate frame. 
 We calculate the tidal tensor using the second-order finite difference across a $3\times3\times3$ cell cube centered on the stellar particle. 
 We then calculate the three eigenvalues of the tidal tensor, $\lambda_1, \lambda_2, \lambda_3$, and use 
 $$ \lambdam \equiv {\rm max}|\lambda_i| $$
 as an upper limit of the tidal strength, following \citet{li_gnedin19}. 
 In the rest of this paper we use $\lambdam$ to refer to the tidal strength.
 %According to the Poisson equation, the trace of the tidal tensor $\lambda_1+\lambda_2+\lambda_3$ equals $4\pi G\rho$.
\rev{According to the Poisson equation, the trace of the tidal tensor $\lambda_1+\lambda_2+\lambda_3$ equals $-4\pi G\rho$.} 
 
 The gravitational potential calculated in a cosmological simulation contains a cosmological term due the universal expansion \citep[e.g.,][]{martel_shapiro98}. We confirmed that this term is much smaller than the Newtonian potential we wish to evaluate around star clusters. It contributes less than 0.4~Gyr$^{-2}$ to $T_{ij}$ at redshifts $z<9$, while the typical Newtonian values are above $10^3$~Gyr$^{-2}$. Therefore, we can differentiate the values of $\Phi$ taken directly from the simulation snapshots.
 
 \begin{table}
  \centering 
  \begin{tabular}{lccccc}
   \toprule
           & \multicolumn{3}{c}{Cell size (kpc)} & $\lambdam$>$10^4$~Gyr$^{-2}$ & age<150~Myr\\
   Run     & 25\% & 50\% & 75\% & 50\% & 50\%\\
   \midrule
   SFE10   & 0.13 & 0.19 & 0.28 & 0.10 & 0.17 \\
   SFE50   & 0.14 & 0.19 & 0.31 & 0.10 & 0.19 \\
   SFE100  & 0.14 & 0.21 & 0.34 & 0.10 & 0.17 \\
   SFE200  & 0.20 & 0.38 & 0.71 & 0.11 & 0.17 \\
   SFE200w & 0.22 & 0.31 & 0.52 & 0.13 & 0.17 \\
   \bottomrule
  \end{tabular}
  \caption{The median and interquartile range of the sizes of cells containing massive star clusters in each run, combined over all analysed snapshots. The last two columns show the median size for a subset of cells with strong tidal field or young stellar particles.}
  \label{tab:cellsize}
 \end{table}

 \begin{figure*}
  \begin{center}
   \includegraphics[width=.49\textwidth]{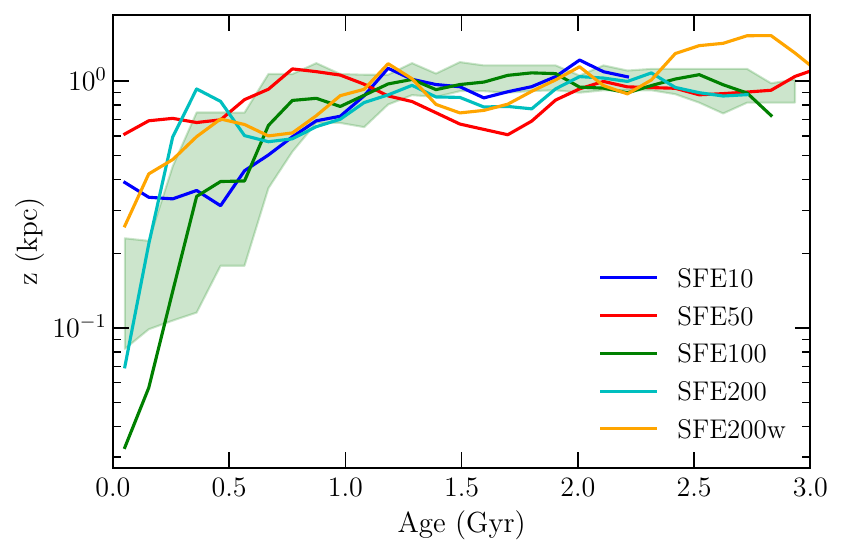}
   \includegraphics[width=.49\textwidth]{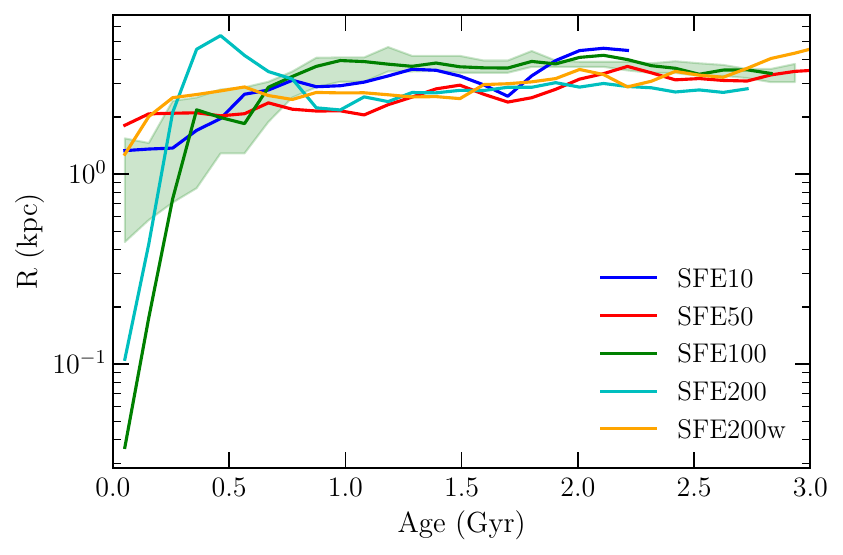}
   \includegraphics[width=.49\textwidth]{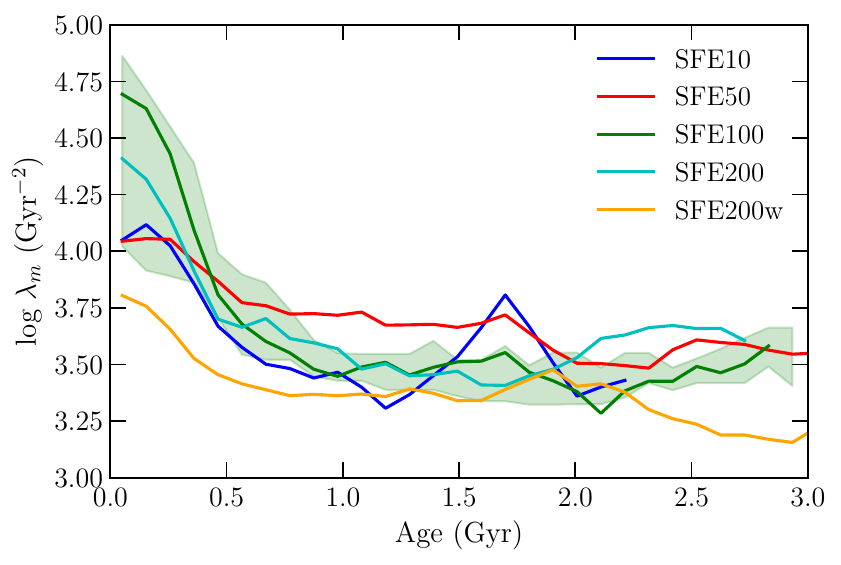}
   \includegraphics[width=.49\textwidth]{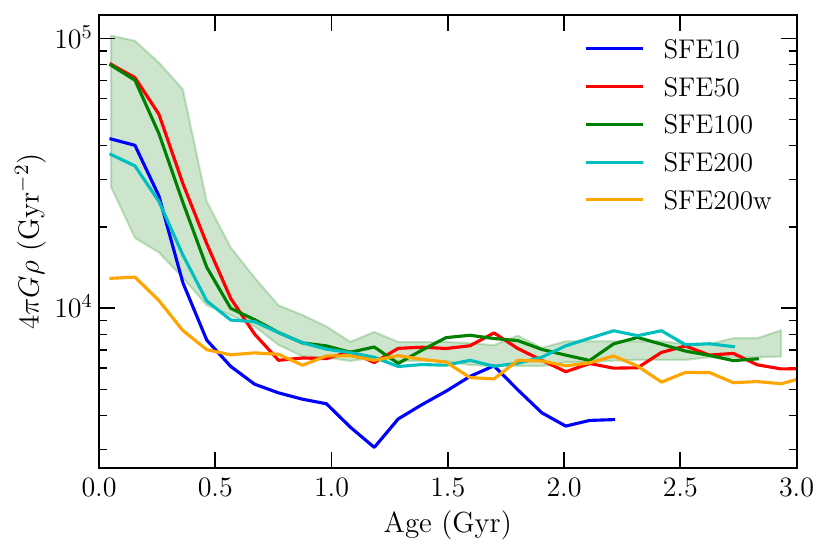}
  \end{center}
  \vspace{-4mm}
  \caption{Median values of \rev{height} $z$, radius $R$, maximum eigenvalue of the tidal tensor $\lambdam$, and ambient density $\rho$ in age bins from the simulation snapshots. Only star clusters born in the main galaxy are included (first found within 7~kpc from the center and younger than 150~Myr). The lines are smoothed with a Savitzky-Golay filter with window size of 900~Myr. Shaded regions are the interquartile range (25-75\%) in a moving window of 700~Myr for the SFE100 run.}
  \label{sum_medevol} 
 \end{figure*}

 We chose the averaging scale of the ambient density of 100~pc because it matches best the scale on which we calculate the tidal tensor.
 The size of cells containing the traced stellar particles at each snapshot ranges from tens of pc to kpc scale, as they appear in various locations in the halo.
 We include the median value and the interquartile range of the cell sizes for each run in Table~\ref{tab:cellsize}.
 The median of the total distribution is $200-300$~pc, while the cells containing clusters younger than 150~Myr have median size about 170~pc. In regions with strongest tidal field ($\lambdam>10^4$~Gyr$^{-2}$) the cells are 100--130~pc.
 Therefore, the averaging scale of the ambient density does not need to be much smaller than that. 
 The 75\% percentile for the SFE200 run is larger than for the other runs because it has a group of stellar particles in a low density region far from the galaxy center. This group biases the overall distribution of cell sizes. 
 We will discuss the scale for the tidal tensor calculation in more detail in Section~\ref{sec:discuss}.
 
\section{Results} \label{sec:result}

\subsection{Overall evolution of cluster location, tidal field, and ambient density} \label{sec:overall}

 First we focus on {\it in situ} star clusters, which remain in the main halo at all times and do not transfer to another halo during mergers. 
 Our criterion for a cluster to be identified as formed {\it in situ} is to be first found within 7~kpc from the main galaxy center and younger than 150~Myr. 
 We chose the 7~kpc threshold to guarantee that the clusters formed in the main galaxy.
 If we change this criterion to 5~kpc, it would only change the identified origin of one cluster. 
 We chose the 150~Myr age threshold because the largest time interval between two consecutive snapshots is about 150~Myr, and we want to ensure that the clusters we found were not formed earlier in other places and then brought in to the main galaxy.
 
 We exclude a group of clusters in the SFE200 run that are likely born in a nearby satellite galaxy at $z\approx 5.2$, although they fit the other {\it in situ} criteria.
 This satellite is at $\sim2.5$~kpc from the main galaxy center at that snapshot, so it would be challenging to revise the criterion to avoid this group of clusters automatically. 
 However, this group has highly elliptical orbits around the main galaxy and never gets as close to the center again, so we do not consider them as born \textit{in situ} and exclude them from Figure~\ref{sum_medevol}. % and \ref{corr_slope}.

 We trace the evolution of the four variables ($z$, $R$, $\rho$, $\lambdam$) for these massive clusters in the simulation snapshots.
 We first describe the overall evolution of these quantities with cluster age. 
 In Figure~\ref{sum_medevol} we show the median values in age bins for clusters born in the main galaxy.
 This plot shows that \textit{in situ} star clusters tend to move away from the center and out of the galaxy disc, into the regions of lower density and weaker tidal field. 
 We can also see that the four tidal variables change the most in the first 500--1000~Myr after cluster formation, and subsequently reach almost constant plateaus. 
 Note that the lines show only the average trend -- most of {\it in situ} clusters (about 70\%) move outward, but some ($\sim$30\%) move inward. 
 Some clusters have elliptical orbits that are on average farther away from the galaxy center than at formation. 
 Since the snapshots are taken at arbitrary times, the clusters can be found at random phases of their orbits. 
 The outward-going trend is thus an average effect of all massive star clusters that are born in the main galaxy.
 
\begin{figure}
   \includegraphics[width=\columnwidth]{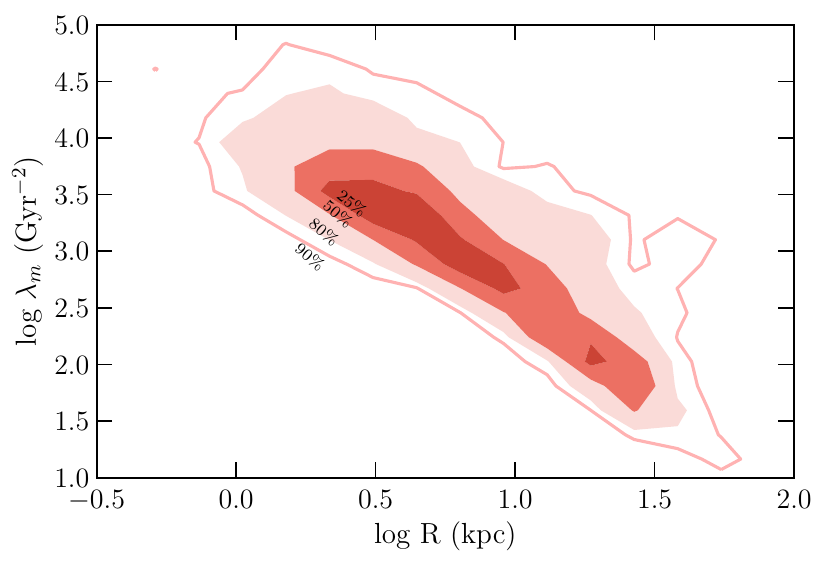}
   \vspace{-5mm}
   \caption{Distribution of $\lambdam$ and $R$ shown as 25\%, 50\%, 80\%, and 90\% contours of all traced massive star clusters. All five simulation runs and all epochs are combined. There is a negative correlation between $\lambdam$ and $R$ with a power law slope of $-1.23$.}
  \label{corr_tide_R} 
 \end{figure}

 Note that the position ($z$, $R$) and tidal field related properties ($\lambdam$, $\rho$) are not independent of each other. 
 Density decreases with distance from the galaxy center and from the disc plane, and tidal strength is positively correlated with density. 
 In Figure~\ref{corr_tide_R} we show a correlation between the galactocentric distance and the tidal field at a given point. 
 Here we include all star clusters (not only {\it in situ} ones)  because this relation describes the overall potential of the main galaxy.
 The distribution extends far into the lower right corner because several clusters move to large galactocentric radii, and thus small $\lambdam$ regions, and some of the {\it ex situ} clusters come from large radii. 
 These star clusters can bias the whole distribution due to a limited number of massive star clusters we trace.
 The farther away from the center, the weaker is the tidal field and the lower is the ambient density. 
 A similar negative correlation holds for $\lambdam$ and the height above the disc plane $z$, but with a different slope from the relation in Figure~\ref{corr_tide_R} and larger scatter. 
 %The values of $\lambdam$ are close to $4\pi G\rho$, as $\lambdam$ is the largest eigenvalue of the tidal tensor. 
 \rev{The values of $\lambdam$ closely correlate with $4\pi G \rho$.}
 This means the decrease in the tidal strength experienced by clusters is usually correlated with the clusters going away from the disc plane, out of the galaxy, and into the lower density regions.
 
 As we can see from Figure~\ref{sum_medevol}, the position and tidal field of the star clusters do not change linearly with time. 
 We can calculate a slope of this evolution, but must be careful on what timescale to calculate it. 
 The slope is steeper and has larger scatter among different clusters shortly after cluster formation. 
 Therefore, for the following analysis we discuss slopes of the quantities over the first 1~Gyr after formation, since this is the interval when their position and tidal field change the most. 
 %Not only the tidal variables is correlated with distance from the galactic center, but their derivatives are also correlated. 
 %Figure~\ref{corr_slope} shows that the correlation between  d$\log\lambdam$/d$\log t$ and d$\log R$/d$\log t$ is similar among all five runs, as is the correlation of $\lambdam$ and $R$. 
 %And the slope of the $d\log\lambdam/d\log t - d\log R/d\log t$ relation is similar to the slope of the $\log\lambdam - \log R$ relation in Figure~\ref{corr_tide_R}.
 %This correlation exists regardless of the time range in which we fit the linear relation.
 %If we fit power-law slopes to the data in Figure~\ref{corr_tide_R} and in Figure~\ref{corr_slope}, we obtain $-1.23$ and $-0.76$, respectively.
 %The accuracy of galaxy center determination is 10~pc, and we use this as the uncertainty in the fitting. 
 %The decrease in the tidal strength is correlated with the star clusters going away from the center.

 %\begin{figure}
 %  \includegraphics[width=\columnwidth]{./figures/sum_slopes_log_tide_R_1Gyr_corr.pdf}
 %  \vspace{-5mm}
 %  \caption{Evolution of power-law slopes of galactocentric  radius and tidal strength for \textit{in situ} clusters. The slopes are calculated in the first 1~Gyr after cluster formation. A linear relation between the points in this figure has a slope of $-0.76$.}
 %  \label{corr_slope} 
 %\end{figure}

\subsection{Time spent in strong tidal fields}

 %The most important statistic relevant to disruption is the amount of time clusters spend in regions of strong tidal field. 
%\rev{Clusters experience the most disruption in strong tidal fields. }
\revv{The disruption processes of clusters are integrated effects over time. Both the intensity of the tidal field and the duration of strong tidal events matter the most in cluster disruption. }
 However, not only the finite spatial resolution makes us underestimate the tidal strength, but the time between our simulation snapshots is also relatively long ($\sim$100--150~Myr). 
 We only have access to the potential at these snapshots, which are taken at random phases of the orbit and likely miss the short periods when the clusters experience strong tidal field.
 
\begin{figure}
   \includegraphics[width=\columnwidth]{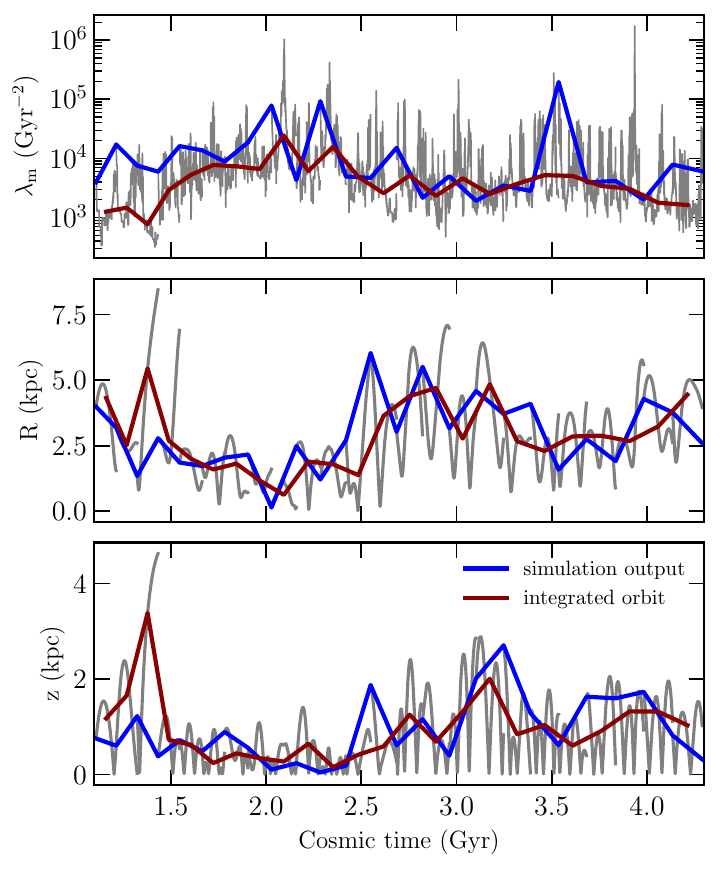}
  \vspace{-5mm}
  \caption{Calculated orbit for one star cluster (gray lines) assuming the potential is not changing between simulation snapshots. Blue lines connect values output at the snapshots. The actual orbit goes through a wide range of distances from the galaxy center and tidal field strengths, so that the snapshot values may not be representative of the tidal history (although the orbit post-processing is definitely inaccurate). Red lines are the medians of the post-processed quantities between the snapshots.}
  \label{hack}
 \end{figure}

 We consider two ways to get around this limitation. 
 We can either reconstruct the orbits between two snapshots using the information at the snapshots, or re-run the simulation and output the tidal field at shorter intervals. 
 Figure~\ref{hack} shows the former, an example of calculating one cluster orbit in the SFE50 run assuming the potential is fixed between two consecutive snapshots.
 We deduct the bulk velocity of the galaxy to account for the moving galaxy center. 
 Every block of gray curves is one orbit calculated between two snapshots.
 The blocks are not smoothly connected because the calculated orbits cannot guarantee to match the points of the next snapshot.
 This figure illustrates that $z$, $R$, and $\lambdam$ vary rapidly and the values taken at the simulation snapshots may not be representative of the actual orbit.
 
 We compared the tidal field values of our calculated detailed orbit with that from the tidal writeout. The values are close within $\sim$100~Myr from the calculation start point, but our calculated orbit could not capture the tidal field if there are encounters with high-density structures. We also tried interpolating the potential between the start and end epochs and it does not seem to improve the accuracy of our calculated orbits. 
 
 We compare the slopes calculated from separate snapshots and detailed orbits and found the slopes from snapshots can basically match slopes from detailed orbits.
 We fit the median of the values and median of the time of these red lines.
 We also connect the values at each snapshot by blue lines, so that each section of the gray curves start on the blue lines, and fit a slope to these data.
 We compared the slopes of these two linear fits and found the two slopes are basically consistent but with a relatively large scatter. 
 %This confirms the legitimacy of our analysis using information from snapshots loosely spaced in time to the analysis in the previous section.
 \rev{This shows that, although using snapshots loosely spaced in time loses information of detailed orbital evolution, it captures the overall \revv{low-frequency} evolution of position and tial field on a larger timescale, as presented in the analysis in the previous section. } 
 
\begin{figure}
   \includegraphics[width=\columnwidth]{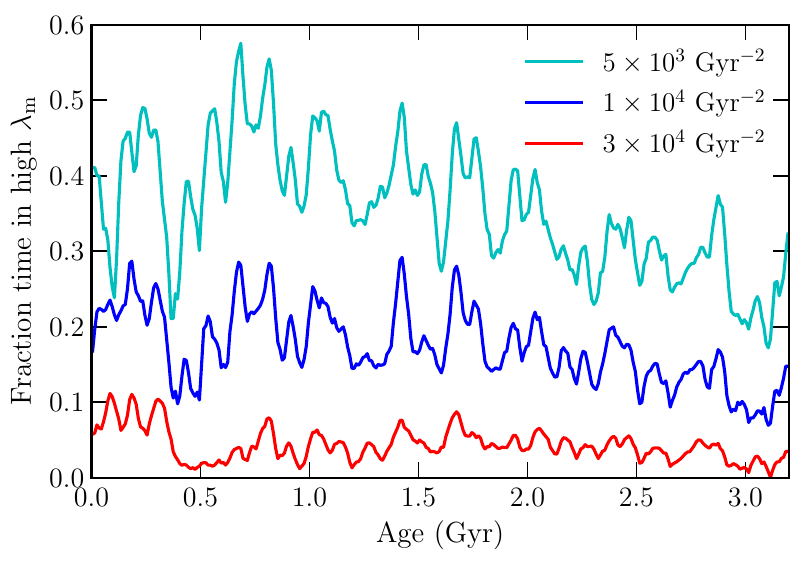}
  \vspace{-5mm}
  \caption{Average fraction of time in strong tidal field above three threshold values ($\lambdam>3\times10^4$~Gyr$^{-2}$, $1\times10^4$~Gyr$^{-2}$, and $5\times10^3$~Gyr$^{-2}$) for 38 star clusters in SFE50 run. The fraction is calculated in bins of 10~Myr from detailed orbits integrated assuming fixed potential between the snapshots. The time fraction in strong tidal field regions is larger at early ages.}
  \label{fracinhack_aver}
 \end{figure}

 %Since the disruption of a star cluster is related to the integrated time it spends under the effects of strong tidal field, we calculate the fraction of the time between every two snapshots that a given cluster experienced tidal field with $\lambdam > 3\times10^4\,{\rm Gyr}^{-2}$.
\rev{Since the disruption of a star cluster is related to the integrated time it spends in tidal field, we calculate the fraction of the time that a given cluster experienced in tidal field with $\lambdam > 3\times10^4$~Gyr$^{-2}$. }
 The choice of this threshold is based on the disruption timescale as well as the evolution of $\lambdam$ with age --  most star clusters have $\lambdam$ below several times $10^4$~Gyr$^{-2}$.
 In Figure~1 of \citet{li_gnedin19}, based on the same simulation suite, $10^4$~Gyr$^{-2}$ is also a fairly high value that is reached in a very small portion of the time a few hundred Myr after the example cluster is formed.
 We then average this time fraction for all \textit{in situ} star clusters in the SFE50 run and plot the result in Figure~\ref{fracinhack_aver}. 
 In this figure we can see that immediately after formation clusters spend more time in regions of high tidal strength. 
 After they grow older and move away from the galaxy center they spend less time in high tidal field regions and experience less disruption. 
 This is consistent with the previously mentioned outward motion of the majority of the {\it in situ} clusters, and the decreasing tidal strength with cluster age.
 
 Our chosen threshold value for $\lambdam$ is somewhat arbitrary, and therefore we also show results for lower threshold values.
 As expected, decreasing the threshold increases the time fraction. 
 When the threshold value decreases to $5\times10^3$~Gyr$^{-2}$, the time fraction rises to $\sim$40--50\% in the beginning. 
 This indicates that this threshold value $5\times10^3$~Gyr$^{-2}$ is a typical value of the tidal strength in the first Gyr after the formation of these massive star clusters. 
 Rather than using the arbitrary time and location given by the widely separated snapshots, with the detailed orbits we can see that even immediately after formation the clusters do not spend all of their time in strong tidal field regions, but only $\sim$10\% of the time. 
 As they grow older, this fraction drops to a few percent.
 
 As an alternative to reconstructing cluster orbits, we have the SFE200w run with detailed output of all massive star clusters at densely spaced intervals of 0.1~Myr.
 We also compared the $R$ and $\lambdam$ slopes from snapshots and from tidal output for this run, to examine if using tidal field and position from snapshots is acceptable. 
 It turns out that the slopes are consistent.

\begin{figure*}
  \begin{center}
   \includegraphics[width=0.9\linewidth]{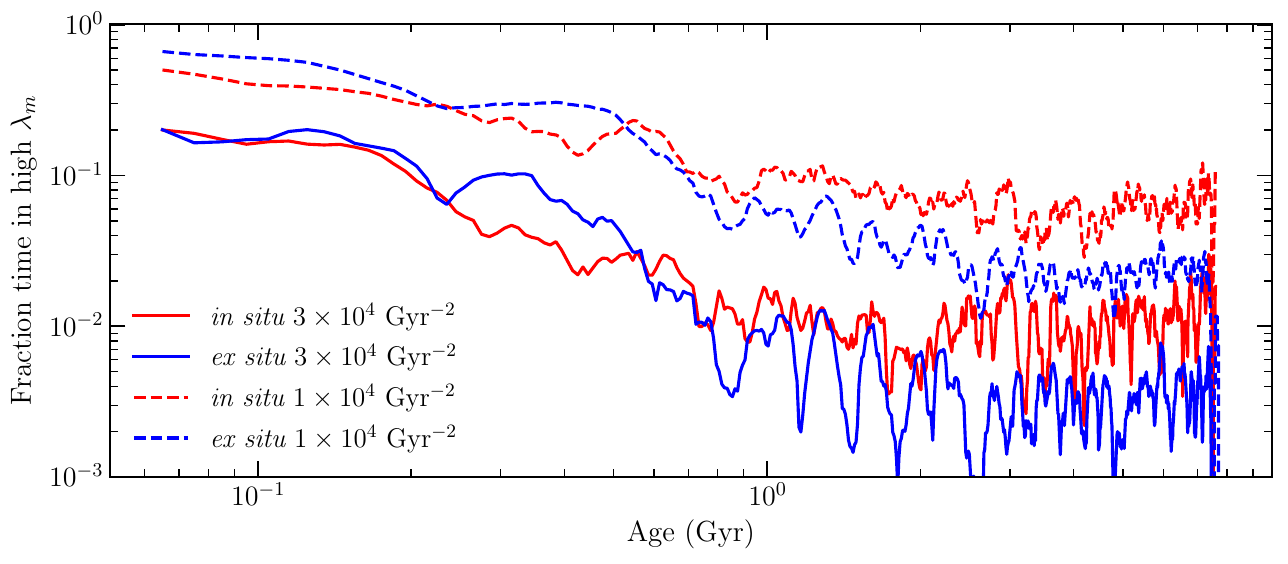}
  \end{center}
  \vspace{-4mm}
  \caption{Average fraction of time in tidal fields with $\lambdam>3\times10^4$ Gyr$^{-2}$ (solid lines) and $\lambdam>1\times10^4$ Gyr$^{-2}$ (dashed lines) in SFE200w run. The fraction is calculated in 10~Myr bins and smoothed with the Savitzky-Golay filter. The average value for all {\it in situ} clusters is shown as red lines, the average for all {\it ex situ} clusters as blue lines. Again, the time fraction in strong tidal field is higher at early ages and decreases with time. {\it Ex situ} star clusters end up farther from the center, in weaker tidal fields.}
  \label{fracinwriteout_aver}
 \end{figure*}

 \rev{We could also calculate the average time in strong tidal field regions for this one run with detailed tidal output. }
 Using this SFE200w run, in Figure~\ref{fracinwriteout_aver} we plot the average time fraction in strong tidal fields for {\it in situ} and {\it ex situ} star clusters. We do not include star clusters that have not yet merged into the main halo by the last available snapshot ($z \approx 0.79$).
 
 The evolution of the time fraction in the SFE200w run is similar to the SFE50 run shown in Figure~\ref{fracinhack_aver}: $\sim 10$--$20\%$ in the beginning and then a drop to a few percent at old ages. 
 The average time fractions in strong tidal field regions for the {\it in situ} and {\it ex situ} clusters are similar in the first few million years after formation but differ significantly at old ages.
 When their host galaxies merge with the main galaxy, the {\it ex situ} clusters have larger total energy than the {\it in situ} clusters, and therefore end up farther away from the galaxy center and in weaker tidal field regions. We discuss this more in the next section. 

\subsection{Starting and ending locations}

\begin{figure}
   \includegraphics[width=\columnwidth]{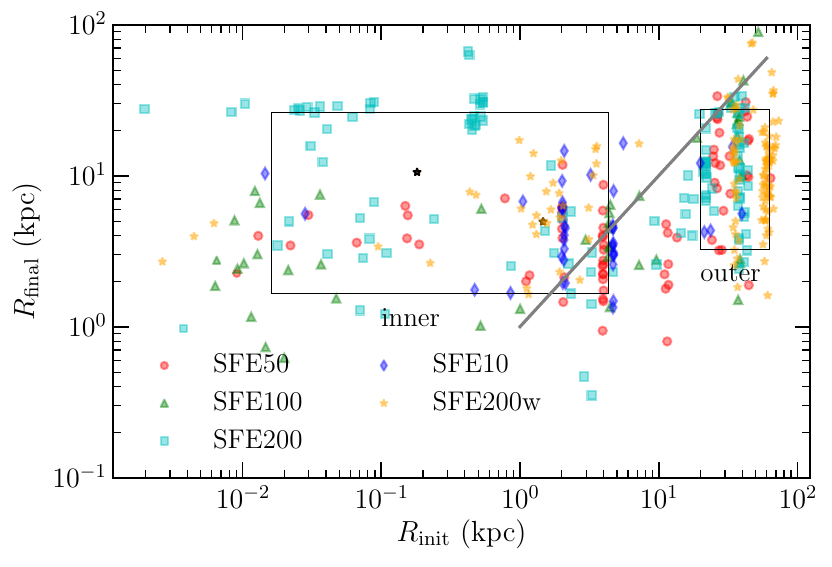}
  \vspace{-5mm}
  \caption{Radius from the center of the main galaxy in the final snapshot vs. radius when the clusters initially appear. Rectangular frames mark the $20\%-80\%$ percentiles of the inner ($R_{\rm init}\leq 7$~kpc) and outer ($R_{\rm init}>7$~kpc) groups of clusters. The gray line marks 1:1 relation. Majority of clusters born in the main galaxy ($\sim$80\%) move outward. (This fraction is different from the 70\% we quote in Section~\ref{sec:overall} because the criterion for {\it in situ} is not exactly the same and we are comparing starting and ending position instead of in 1~Gyr.) On the other hand, most of clusters brought in by mergers (outer group) move inward. Nevertheless, majority of the {\it in situ} clusters still end up closer to the galaxy center then the {\it ex situ} clusters.}
  \label{RiRf}
 \end{figure}

 Star clusters have different tidal histories depending on where they were born.
 \rev{This difference could reflect in different expectations of final bound fraction for star clusters formed in the main galaxy or satellites, and also scatter in the bound fraction for clusters from the same origin. }
 Inspired by Figure~\ref{fracinwriteout_aver}, we examine the evolution of cluster location in Figure~\ref{RiRf}. 
 If the initial distance from the galaxy center $R_{\rm init}$ is small, then the cluster is most likely born {\it in situ}, and vice versa.
 We divide all star clusters into the inner ($R_{\rm init}\leq7$~kpc) and outer ($R_{\rm init}>7$~kpc) groups to roughly distinguish {\it in situ} and {\it ex situ}.
 Note that unlike Section~\ref{sec:overall}, here we do not exclude the group of clusters that formed in a satellite galaxy near the pericenter of its orbit around the main galaxy.
 
 We find that about 80\% of the {\it in situ} clusters move outwards from their birth locations. 
 At the final snapshot, most {\it in situ} clusters still end up within $\sim$7-8~kpc of the galaxy center.
 As expected, most of the {\it ex situ} clusters end up much farther from the galaxy center. 
 {\it Ex situ} clusters coming from the same origin most likely form a stream and have similar orbits in the main halo. In a given snapshot, they may be at different orbital phases but all of them can reach the furthest point of its group.
  
\begin{figure*}
   \includegraphics[width=0.9\linewidth]{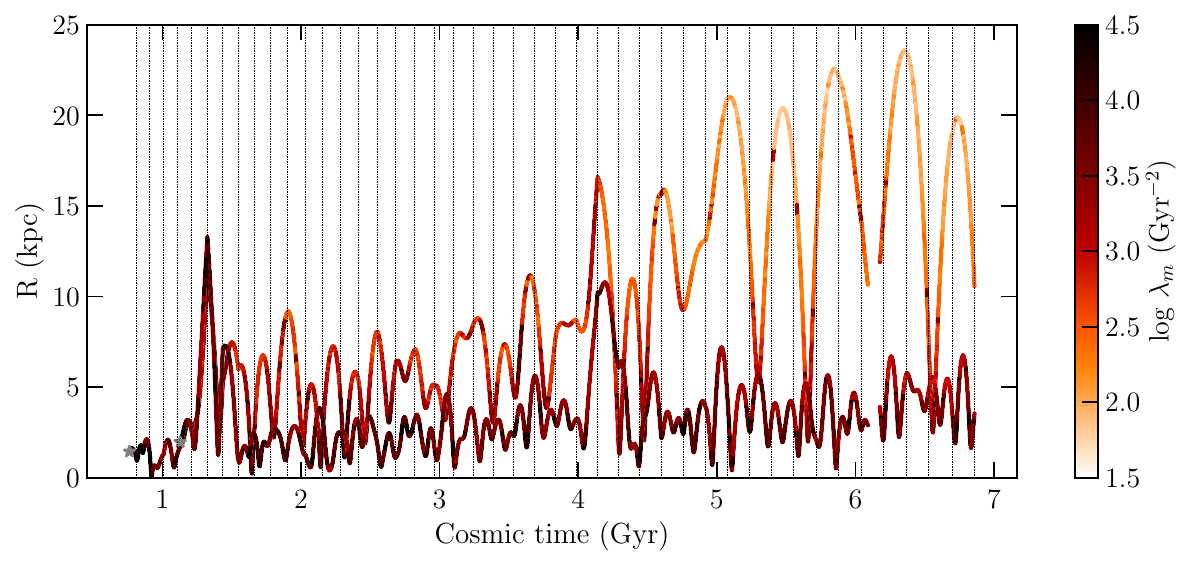}
   \vspace{-2mm}
   \caption{Examples of the orbits and $\lambdam$ of two clusters in the SFE200w run. One stays in the main galaxy, one goes to a higher energy orbit. The gray stars are the starting points (around 15~Myr of age) of the two curves.}
   % cmr.ember https://cmasher.readthedocs.io/user/cmap_overviews/mpl_cmaps.html
   \label{orb_eg}
 \end{figure*}

 In Figure~\ref{orb_eg} we show examples of the orbits of two {\it in situ} clusters in the SFE200w run, corresponding to the orange stars with black edges in Figure~\ref{RiRf}.
 One cluster stays deep in the main halo throughout the whole simulation time.
 The other escapes the main galaxy on a highly elliptical orbit after the energy jumps which usually happen during galactic mergers.
 The orbits are color-coded by the tidal strength $\lambdam$.
 The inner cluster remains in the regions of relatively strong tidal field throughout the simulation time, while the escaping cluster experiences weaker tidal field most of the time.
 Note that the times of output snapshots do not coincide with the timing of strong tidal field, especially for highly elliptical orbits as clusters spend most of the time at larger radii. 
 \rev{From this example we can see that even star clusters formed in the same galaxy could have different orbits and therefore different tidal histories, and the final survival of these clusters. }
 
\begin{figure}
   \includegraphics[width=\columnwidth]{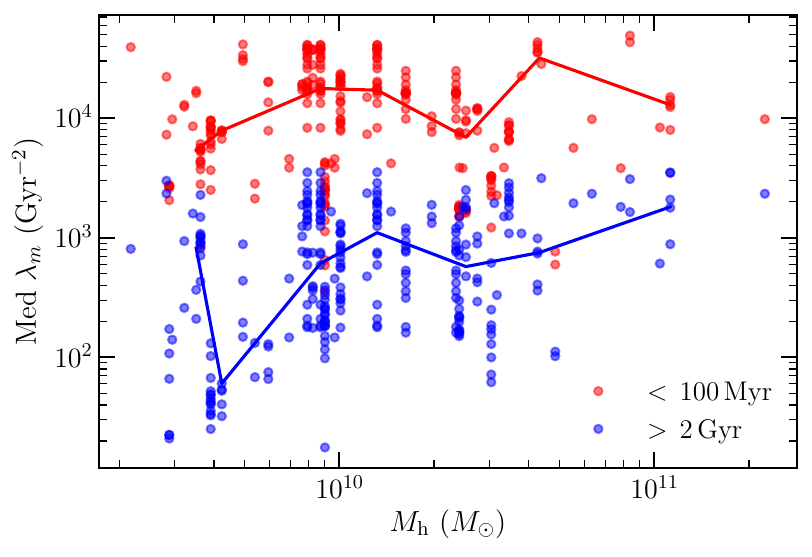}
  \vspace{-5mm}
  \caption{Median values of $\lambdam$ over two time periods: right after cluster formation (red) and at late times (blue) for star clusters in the SFE200w run, vs. host halo mass at formation. Each point represents one cluster. The lines are the median of the corresponding points. There is no noticeable dependence of the tidal field at birth on the host mass. In contrast, the final tidal field strength increases with host mass. Clusters born in smaller halos tend to end up in weaker tidal field regions, possibly because when they merge into the main halo they end up farther from the center.} 
  \label{lambda_m_Mhalo}
 \end{figure}

 We can also examine the difference in the tidal field experienced by {\it in situ} and {\it ex situ} clusters by looking at the host galaxy mass. 
 In Figure~\ref{lambda_m_Mhalo} we not only include clusters that end up in the main halo, but also take a few massive star clusters in all halos and record their host halo mass at the time of formation.
 Then we compare the tidal field strengths before the star clusters are 100~Myr old and after they are 2~Gyr old, which we call `initial' and `final' values.
 Unlike the previous plots, here we do not use any threshold on $\lambdam$.
 The initial tidal strength does not differ much for clusters formed in smaller or larger host galaxies, but there does seem to be a difference in the final tidal field. 
 This is consistent with what we found before -- clusters that form in smaller galaxies (which will potentially merge with the main galaxy) end up in less dense regions, and farther away from the galaxy center.
 
 \subsection{The bound fraction}

 We can now look at the bound fraction of the star clusters we trace. 
 Since star clusters lose mass due to a variety of processes, we write the fraction of the initial cluster particle mass that remains gravitationally bound at age $t$ as
 \begin{equation}
  f_{\rm bound}(t) = f_i\, f_{\rm se}(t)\, f_{\rm dyn}(t)
 \end{equation} 
 where $f_i$ is the initial bound fraction accounting for gas expulsion during cluster formation stage, $f_{\rm se}$ is accounting for mass loss due to stellar evolution, and $f_{\rm dyn}$ is accounting for tidal stripping of stars.
\rev{The initial bound fraction $f_i$ the fraction of mass in a star-forming complex that remains bound in the early phase, calculated as
$$ f_i \equiv {\rm min}\left(\frac{\epsilon_{\rm int}}{\epsilon_{\rm core}}, 1\right)$$
where $\epsilon_{\rm core}=0.5$ is the correction factor for mass loss due to protostellar outflows,
and $\epsilon_{\rm int}$ is the integral star formation efficiency -- the mass of the active stellar particle divided by the maximum baryon mass of the GMC throughout the whole course of cluster accretion (see \citet{Li:2018aa} for more details). }
 The fraction $f_{\rm dyn}$ is calculated in run time using the current tidal tensor for each star cluster \citep{li_gnedin19}.
 \rev{Since here we focus on disruption by tidal process, i.e. $f_{\rm dyn}$, for simplicity we use the phrase ``bound fraction" to represent $f_{\rm dyn}$ in the following text. }
 
 \begin{figure*}
   \includegraphics[width=.49\textwidth]{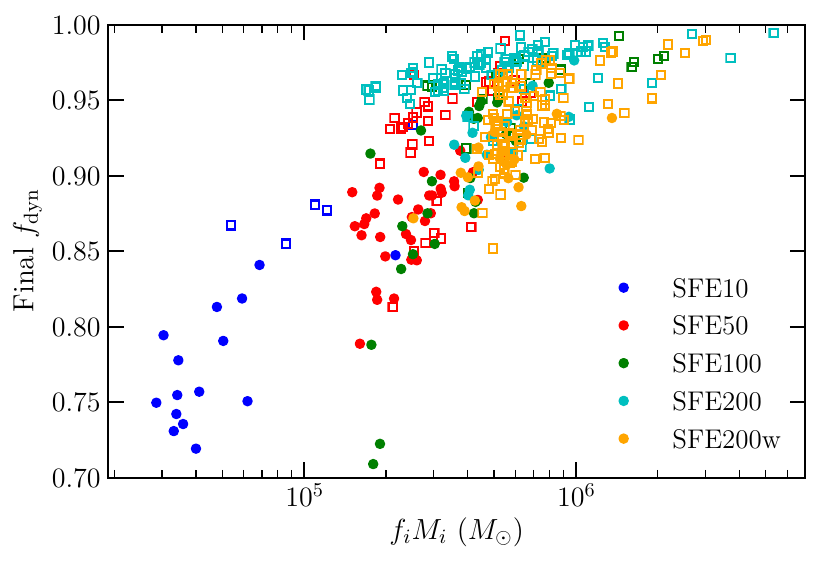}
   \includegraphics[width=.49\textwidth]{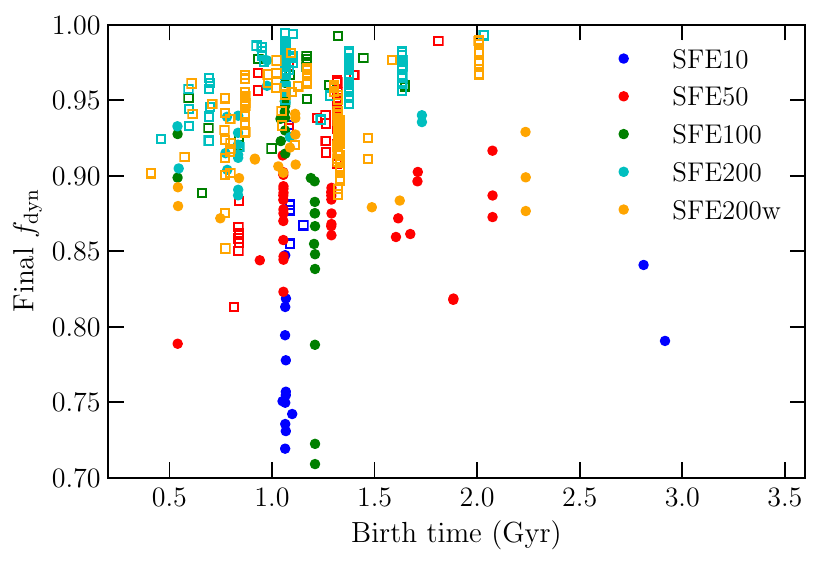}
  \vspace{-2mm}
  \caption{Relation between the bound fraction of clusters at last available snapshot and their initial mass (\textit{left panel}) or birth time (\textit{right panel}). Different colours represent different runs. Solid circles are for \textit{in situ} clusters, open squares are for \textit{ex situ} clusters. The final bound fraction positively correlates with both the initial mass and birth time. Star clusters born in the main galaxy generally end up with smaller bound fractions than those born in satellite galaxies.} 
 \label{sum_fb_Mi}
 \end{figure*}
 
 The timescale for tidal disruption can be approximated as
 \begin{equation}
    t_{\rm tid}(M,\,\Omega_{\rm tid}) = 10\,{\rm Gyr} \left(\frac{M(t)}{2\times10^5\Msun}\right)^{2/3} \frac{100\,{\rm Gyr}^{-1}}{\Omega_{\rm tid}(t)}.
    \label{eq:ttid}
 \end{equation}
 where $M(t)$ is the current cluster mass, and the frequency $\Omega_{\rm tid}(t)$ is related to the maximum eigenvalue of the tidal tensor along the cluster trajectory as
 $$ {\rm \Omega_{tid}^2} = \frac{1}{3}\lambdam $$
 \citep{li_gnedin19}.
 \revvv{This is a simplification of the actual disruption calculation that must include the full tidal field and the structure of the cluster. We use it here for convenience to obtain first estimates of the cluster mass loss. We compare the results with an alternative simplified expression for $t_{\rm tid}$ in Appendix~\ref{appendix2}. We find general consistency between the two methods, however note that neither expression can give an accurate value for the actual mass loss.}

 During one time step from time $t_n$ to $t_{n}+{\rm d}t$, the bound fraction $f_{\rm dyn}$ was calculated as 
 $$
    f_{n+1} = \exp\left(-\frac{{\rm d}t}{t_{\rm tid}^{\rm out}}\right) \, f_{n}.
 $$ 
 However the disruption timescale $t_{\rm tid}^{\rm out}$ used in these simulations had the normalization a factor 2.5 shorter than our current estimate given by Equation~\ref{eq:ttid}, and also because of a coding mistake, the value of $\lambdam$ was overestimated by the expansion scale factor $a_{\rm box}(t)$. 
 \rev{We correct for this mistake using a method described in Appendix~\ref{appendix}.} 

 \subsection{Testing the tidal disruption timescale}

 Tidal disruption is an important component of semi-analytical models of globular cluster evolution. 
 \rev{Analytical models usually use a single parameter to describe tidal strength. }
 For example, \citet{choksi_etal18} introduced a parameter to normalize the strength of average tidal field along cluster trajectories. %In our notation,
 %\begin{equation} 
 %   P \equiv \frac{100\,{\rm Gyr}^{-1}}{\Omega_{\rm tid}},
 %   \label{eq:p}
 %\end{equation}
%\rev{where $\Omega_{\rm tid}$ is related to the tidal disruption timescale $t_{\rm tid}$ by Equation~\ref{eq:ttid}. }
 \citet{choksi_etal18} found that the average tidal field $\Omega_{\rm tid}=200\,{\rm Gyr}^{-1}$ matches best the observed mass function of globular clusters systems in nearby galaxies. Our results allow us to calibrate the value of the average tidal field.
 
 From the models of tidal disruption, we expect the cluster bound fraction to depend on cluster age $t$ and initial mass $M_{\rm init}$. If a cluster experiences constant tidal field throughout its lifetime, then the bound fraction can be derived by integrating Eq.~\ref{eq:ttid}. The expected value is
 \begin{equation}
  \tilde{f}_{\rm dyn}(t) = \left[ 1 - \frac{2}{3}\frac{t}{10\,\mathrm{Gyr}} \frac{\Omega_{\rm tid}}{100{\rm Gyr}^{-1}} \left(\frac{M_{\rm init}}{2\times10^5\Msun}\right)^{-2/3} \right]^{3/2}.
  \label{eq:fdyn}
 \end{equation}

 According to this solution, clusters with larger initial mass will have larger bound fraction after the same time. 
 Our higher SFE runs form star clusters with larger initial mass than lower SFE runs, and therefore should have more clusters with large final bound fraction. 
 Apart from $M_{\rm init}$, tidal history also plays a role:  clusters born in the main galaxy should have smaller final bound fraction than clusters born in satellite galaxies and brought to the main galaxy by mergers. 
 
 Figure~\ref{sum_fb_Mi} shows that these general trends are present in our simulations. From the left panel we can see that star clusters with larger initial mass tend to have larger final bound fraction, while from the right panel we can see that clusters formed later tend to have larger bound fraction.
 Since the cluster initial mass correlates with SFE, there is also a correlation between SFE and the final bound fraction: star clusters in runs with lower SFE have lower $f_{\rm dyn}$. 
 
 We also distinguish between {\it in situ} and {\it ex situ} clusters, using the same criteria as in Section~\ref{sec:overall}. 
 We consider the group of clusters that were likely born in a close satellite at $z\approx5.2$ in the SFE200 run as {\it ex situ}, just as in Figure~\ref{sum_medevol}. 
 We can see that the {\it ex situ} clusters have higher final bound fraction than the {\it in situ} ones, indicating that they experience different tidal history, and that the {\it ex situ} clusters on average experience weaker tidal field. 
 This can be seen more clearly in the right panel, where is a quite obvious distinction between the {\it in situ} and {\it ex situ} samples.
 
 Different tidal history for clusters from different origins (host galaxies) has a noticeable effect on their final bound fraction.
 Although the tidal strength immediately after formation is similarly high for clusters born in halos of all mass, the tidal field differs more at late times and affects the final bound fraction.
 Clusters formed in smaller satellite galaxies, which tend to end up farther away from the main galaxy, will have larger final bound fraction than clusters formed in larger galaxies.
 
 \begin{figure}
   \includegraphics[width=\columnwidth]{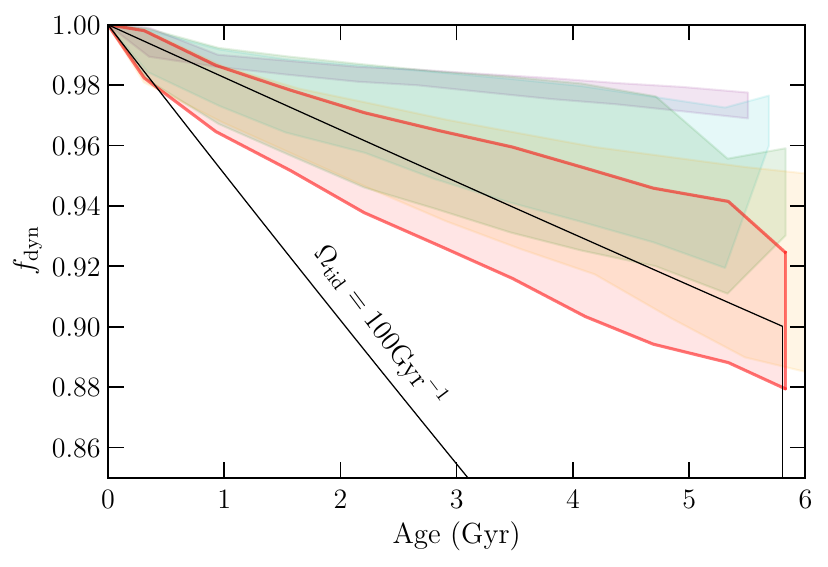}
  \vspace{-5mm}
  \caption{Another illustration of the effect of different tidal history on the final bound fraction. Each shaded region shows the 10-90\% percentiles of the distribution of star cluster masses formed in one galaxy. Red colour is for the main galaxy, emphasized by the red frame; the orange, cyan, and green colours are for different satellite galaxies that merge into the main galaxy, ranked from early to late times. The purple band is for two clusters that originally formed in a satellite of the host galaxy of the cyan clusters and merged into the main galaxy along with the cyan clusters. The bound fraction for clusters from satellites usually decreases more slowly than for clusters formed in the main halo. For comparison, we include a triangular frame that corresponds to the evolution of the bound fraction in a constant tidal field $\Omega_{\rm tid}=100\,{\rm Gyr}^{-1}$, for the same range of initial cluster masses.}
  \label{tidal_writeout_main_fdyn_age}
 \end{figure}

 Figure~\ref{tidal_writeout_main_fdyn_age} shows the evolution of the bound fraction with age for clusters from different origins. Different initial masses of the clusters result in a spread of values of $f_{\rm dyn}$. In general, the bound fraction for clusters from satellites decreases more slowly  than for clusters formed in the main halo.
 
\begin{figure}
  \includegraphics[width=\columnwidth]{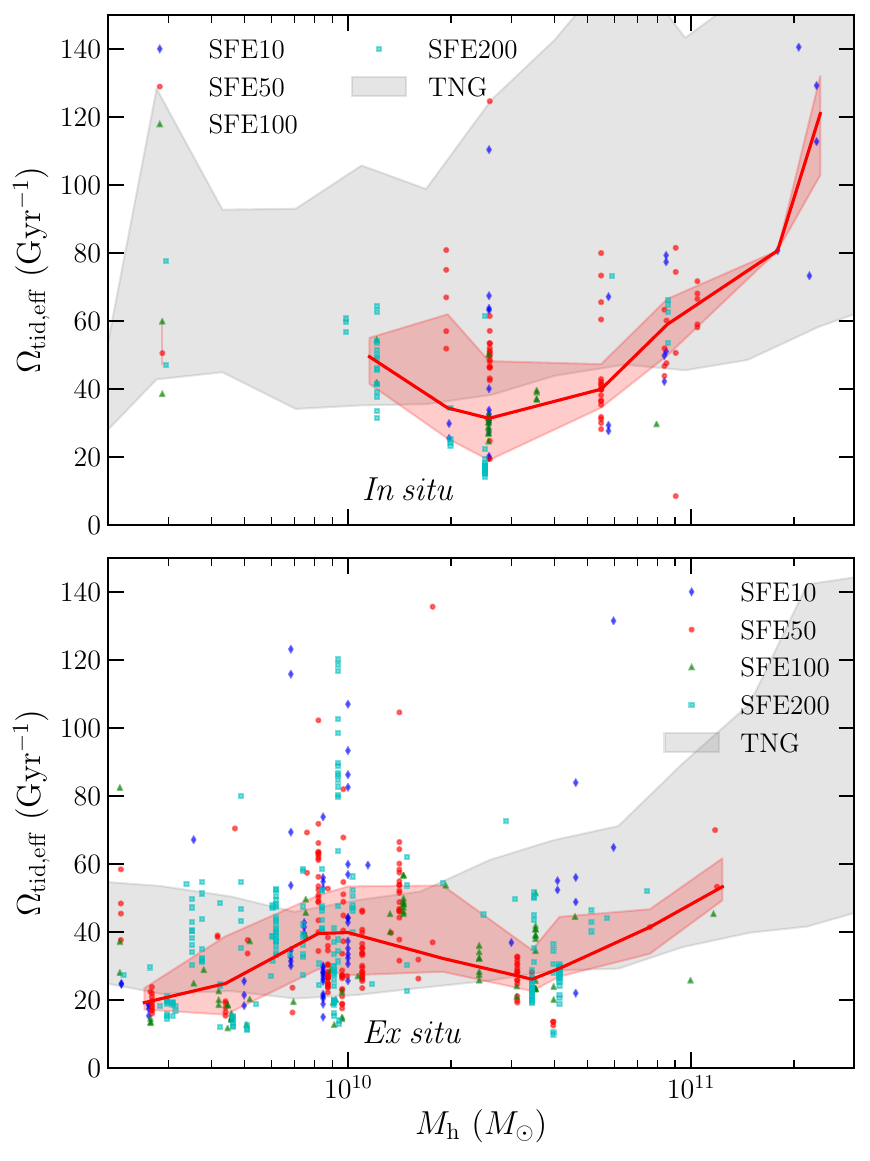}
  \vspace{-5mm}
  \caption{The expected effective $\Omega_{\rm tid,eff}$ value for each cluster versus host halo mass at time of formation, split into {\it in situ} and {\it ex situ} clusters. Different colours represent different runs. The red line and red shaded region are the median and IQR of all the points in bins of $\log{M_{\rm h}}$. {\it In situ} clusters in general experience stronger tidal field than {\it ex situ} clusters. There is a increasing trend with the host halo mass at birth for {\it ex situ} clusters, which means that on average clusters formed in more massive halos experience stronger tidal field. The gray shaded region is the IQR of the $\Omega_{\rm tid,eff}$ value before $z=1.5$ for star clusters with initial mass larger than $2\times10^5\Msun$ modeled by \citet{cyt} based on the TNG50 simulation outputs. The $\Omega_{\rm tid,eff}$ value in the TNG50 simulation has a increasing trend with host halo mass at birth, consistent with our results.}
  \label{sum_P_Mh}
 \end{figure}

 The rate of decrease of $f_{\rm dyn}$ generally slows down with cluster age. This differs from the expectation of Equation~\ref{eq:fdyn} shown as a triangular frame, because the tidal field strength along cluster orbits cannot be treated as a constant. To relate our results to the formulation of Equation~\ref{eq:fdyn} we define the `effective' value of $\Omega_{\rm tid}$ that would produce the actual bound fraction $f_{\rm dyn}$  of the simulated clusters: 
 \begin{equation} \label{Pbar}
  \Omega_{\rm tid,eff} = \frac{3}{2} \frac{10\,\mathrm{Gyr}}{t}
   \left(\frac{M_{\rm init}}{2\times10^5\Msun}\right)^{2/3} 
   (1-f_{\rm dyn}^{2/3}) \times 100\,{\rm Gyr}^{-1}
 \end{equation} 
 and plot $\Omega_{\rm tid,eff}$ in Figure~\ref{sum_P_Mh}. The figure shows significant scatter of $\Omega_{\rm tid,eff}$ for all host galaxies, which reflect a variety of orbital histories in every galaxy. We split the figure into two panels for {\it in situ} and {\it ex situ} clusters. {\it In situ} clusters in general experience stronger tidal field than {\it ex situ} clusters. At the same time, we can see that the average $\Omega_{\rm eff,tid}$ value for {\it ex situ} clusters increases with the host mass, which means that star clusters formed in more massive halos tend to end up with smaller bound fraction than those formed in less massive halos.

 We compare our results with the GC formation model in \citet{cyt}, which is based on the outputs of the TNG50 simulation. 
 In their model, star particles or dark matter particles in the TNG50 simulation are selected to represent star clusters that would form GC systems. 
 They calculated the tidal field surrounding these clusters using the simulation outputs.
 Here we calculated the $\Omega_{\rm tid,eff}$ for each star cluster with the same definition as Equation~\ref{Pbar} and plotted the IQR as a gray shaded region in Figure~\ref{sum_P_Mh}. 
 We used only clusters with initial mass above $2\times10^5\Msun$ to make the cluster sample consistent with our clusters in $f_i M_i$. 
 The $\Omega_{\rm tid,eff}$ values in the TNG50 outputs are similar to those in our simulations, and also show a decreasing trend with the host halo mass at birth, consistent with our results.
 
\section{Discussion} \label{sec:discuss} 

\subsection{Spatial resolution and tidal strength calculation}

 The calculated values of the tidal tensor depend on the spatial scale (cell size). Tidal field calculated on larger scale is typically weaker, as the limited resolution of the simulations may underestimate the actual tidal strength. A typical scale on which we calculate the tidal field is several hundred pc. The median size of cells where the traced star clusters are located is between 100 and 300~pc (see Table~\ref{tab:cellsize}), and the differentiating scale is twice that. The whole range of cells containing the star clusters extends from as small as $\sim$20~pc to as large as 1~kpc in size, depending on the location in the galaxy. This is significantly larger than the typical cluster size, so the actual tidal field experienced by the star clusters is likely underestimated. 

 We cannot increase the simulation resolution but we can reduce it. We checked that if we double the scale on which we calculate $\lambdam$ by combining two cells, the resulting value of $\lambdam$ can decrease. The decrease is $\sim$10\% if the original scale is $\sim$kpc, but it can be as much as a factor of 10 if the original scale is $\sim$100~pc. The amount of decrease scales with the local density because the runtime cell refinement is done according to the matter density. Thus regions of higher density are more affected by the loss of resolution.

\subsection{Comparison with other studies}

 \citet{Renaud:2017aa} performed a cosmological zoom-in simulation of a Milky Way-like galaxy and selected particles older than 10~Gyr to represent globular clusters. 
 %They calculated the tidal tensor and its eigenvalues along the orbits of these particles from the epoch of identification ($z\approx 1.8$) until redshift $z=0.5$. 
 \rev{They calculated the tidal tensor and its eigenvalues along the orbits since redshift $z\gtrsim1.8$ until $z=0.5$. } 
 In agreement with our results, they found that the tidal field experienced by accreted clusters is weaker than that of the {\it in situ} clusters. Also in agreement, before accretion clusters experience generally similar tides regardless of in which galaxy they formed. However, most of their {\it ex situ} clusters are accreted after $z=1.2$, later than the epochs we can probe with our simulations ($z>1.5$). \citet{Renaud:2017aa} do not distinguish between {\it in situ} and {\it ex situ} clusters that merged into the main galaxy during the "rapid increase phase" at redshifts $2\lesssim z\lesssim5$. Although these early accreted clusters are likely to mix better with the {\it in situ} ones than the later accreted clusters, at the end of our simulations they are still located farther from the center and experience weaker tidal field than the true {\it in situ} clusters.
 
 \citet{Renaud:2017aa} also found that the tidal field strength, characterized by the maximum eigenvalue $\lambda_1$ of the tidal tensor, increases with cosmic time after $z \approx 1.5$. This result applies to the later evolution of the overall tidal field and does not conflict with our conclusion that youngest clusters experience the largest $\lambdam$ near their birthplaces. Because of the different time intervals and different measures of the tidal field, the two studies are thus complimentary to each other. 
 Although direct quantitative comparison is not possible, we tried to compare our measures of the tidal field around $z \approx 1.5$. Their average maximum eigenvalue $\lambda_1 \sim 320$~Gyr$^{-2}$ is smaller than our typical $\lambdam>10^3$~Gyr$^{-2}$ around star clusters at late times. However, these two eigenvalues are likely different: $\lambda_1$ is the maximum of the three eigenvalues (and always positive), while $\lambdam$ is the maximum of the absolute value of the eigenvalues and may sometimes take value of $-\lambda_3$. Indeed, Figure~12 of \citet{Renaud:2017aa} shows the average of the other two eigenvalues $\lambda_2$ and $\lambda_3$ is negative and 1-2 times larger than $\lambda_1$. Even taking this into consideration, their average $\lambda_1$ is an order of magnitude smaller than our $\lambdam$. Our simulations have slightly better numerical resolution -- a factor of 2 smaller dark matter particle mass, which may partly explain the difference but not all of it. It is possible that accurate force resolution close to the time of cluster formation is required to capture the earliest and most violent part of their tidal history.
 
 \citet{Kruijssen:2011aa} and \citet{Kruijssen:2012aa} explored star cluster formation and destruction in the context of an isolated galaxy as well as galaxy merger.
 They found that dynamical heating of stellar clusters by tidal shocks can be an order of magnitude higher in interacting galaxies than in isolated galaxies. 
 Although our high-redshift galaxies are more complicated and turbulent than isolated galaxies, we similarly found increased tidal strength experienced by star clusters during galaxy mergers.
 These authors were the first to propose that young clusters that evolve into old globular clusters are those that escape the merging system toward large distances early. They also found low-mass clusters being disrupted preferentially, which is consistent with our work and other current models \citep[e.g.][]{Elmegreen:2010aa,Gieles:2016aa}. 
 \rev{\citet{renaud_gieles13} found in N-body simulations of star clusters that the galaxy major merger only indirectly affects the evolution of clusters by modifying their orbits in or around the galaxies, which is similar in essence to the difference in the bound fraction of {\it in situ} and {\it ex situ} clusters. }

\subsection{On recovering the halo assembly history}

 Recent kinematic studies of the Galactic GC population, coupled with the improved measurements of the age-metallicity distribution, have been used to uncover the assembly history of the Milky Way \citep[e.g.][]{Massari:2019aa,Kruijssen:2019aa}. 
 \citet{Piatti:2019aa} pointed out differences in orbits and kinematics between the \textit{in situ} and accreted Galactic GCs. 
 \citet{Armstrong:2021aa} found a clear kinematic distinction between the metal-rich and metal-poor GCs using Gaia data. 
 Although our simulations extend only to $z=1.5$ and cannot be directly used to reconstruct the assembly history, we also found distinct orbital energy distributions for the {\it in situ} and {\it ex situ} clusters at the last available snapshot.
 Combined with the weaker tidal field experienced by {\it ex situ} clusters after accretion onto the main galaxy, we can expect that a larger fraction of the {\it ex situ} star clusters would survive until the present than of the {\it in situ} clusters.
 
 Other simulations find similar results. 
 \citet{Keller:2020aa} found that more massive halos subject their proto-GCs to stronger disruption in the E-MOSAICS simulations. 
 Their \textit{ex situ} proto-GCs have higher survival fraction than the \textit{in situ} ones, in agreement with our conclusions. 
 They also found that GCs formed in major mergers are more likely to survive than those formed in minor mergers. 
 They argue for the importance of major mergers because they are effective in ejecting clusters to orbits of higher energy and weaker tidal field.
 
 When considering the different tidal field experienced by \textit{in situ} and \textit{ex situ} clusters in models of GC formation and evolution, 
 we note that the normalization of the disruption time ($\Omega_{\rm tid,eff}$) in our Figure~\ref{sum_P_Mh} is averaged not to $z=0$, but only up to $z\sim1.5$. 
 A better calibration of $\Omega_{\rm tid,eff}$ would require simulations that run to redshift zero.

 The orbital information of the GCs can potentially be used to reconstruct the assembly history of the MW.  
 \citet{Pfeffer:2020aa} pointed out that the orbits of GCs 
 are sensitive to the masses and merger timing of their host satellite galaxies. 
 Combined with the results from \citet{Keller:2020aa} and our results 
 that \textit{ex situ} clusters are usually on higher orbits and 
 more likely to survive than \textit{in situ} ones, it is possible 
 to use this information to recover the origins of GCs in the MW. 
 \citet{Kruijssen:2019aa,Kruijssen:2020aa} used the metallicity 
 plus orbital information to recover the merger history of the MW 
 and added a Kraken galaxy corresponding to the `low-energy' GCs, 
 merged prior to the previously found 
 \textit{Gaia}-Enceladus, the Helmi streams, the Sequoia galaxy, 
 and the Sagittarius dwarf galaxy. 
 We caution that using {\it only} the orbital information to 
 reconstruct the merging history may not be effective, 
 because the chaotic merging could mix spatial 
 and kinematic distribution of GCs \citep{Keller:2020aa}. 
 For example, the integrals of motion for our star clusters at the last available snapshot cannot easily distinguish between clusters from different progenitors. 
 In addition, \citet{Garrow:2020aa} argued that the presence of dwarf galaxies 
 can alter the orbital energies and actions of GCs, 
 and that outer clusters are more strongly affected than inner clusters.

\subsection{On parametrization of tidal strength}

 \revvv{Recent literature has several approximations to parametrize strength of the tidal field. One is the largest absolute value of the eigenvalues of the tidal tensor, $\lambdam$, which we used in our previous work. Another is the largest effective eigenvalue  $\lambda_{\rm 1,e} \approx \lambda_1-0.5(\lambda_2+\lambda_3)$ proposed by \citet{pfeffer_etal18} based on numerical experiments with a Plummer sphere potential, and later adopted by \citet{Rodriguez:2022aa}.}
 \revv{We repeated our analysis with $\lambda_{\rm 1,e}$ 
 as an alternative parametrization of the tidal strength 
 to compare the conclusions with our default choice $\lambdam$. 
 We found the value of $\lambda_{\rm 1,e}$ 
 to be within a factor of 2 from $\lambdam$. 
 Clusters experience stronger tidal field 
 and end up with smaller bound fraction using this alternative definition, 
 but our conclusions that tidal strength overall decreases with cluster age 
 and that {\it ex situ} clusters usually end up with larger bound fraction 
 than {\it in situ} clusters remain qualitatively the same. 
 We describe the detailed comparison in Appendix~\ref{appendix2}. 
 }

\section{Conclusions} \label{sec:conclude}

 We use a suite of cosmological simulations with the continuous cluster formation algorithm to investigate the evolution of the tidal field around massive star clusters.
 We calculate the tidal tensor for the simulation snapshots and reconstruct cluster orbits between the snapshots.
 Our main results are summarized below:
 \begin{itemize}
  \item Massive star clusters experience the strongest tidal field in the first 1~Gyr after formation in high-density regions close to their birthplaces. The maximum eigenvalue of the tidal tensor can reach $10^4-10^5$~Gyr$^{-2}$. After about 1~Gyr the median value plateaus at $\lambdam \sim 3\times10^3$~Gyr$^{-2}$.

  \item Most of the {\it in situ} clusters in our simulations 
  move out of the disc plane and to larger distances from the galaxy center (Figure~\ref{sum_medevol}).

  \item The tidal field and cluster positions always correlate. The farther from the center and the galactic disc, the weaker the tidal field (Figure~\ref{corr_tide_R}). A typical relation is $\lambdam \propto R^{-1.2}$. %The rates of change with time of the tidal strength and galactocentric radius correlate similarly (Figure~\ref{corr_slope}).

  \item Cluster orbits are typically eccentric, alternating between regions of high and low density. Simulation snapshots record  only random phases of the orbit, which may not capture the regions of strongest tidal field. Accurate calculation of tidal disruption requires detailed cluster orbits at time intervals $\lesssim 1$~Myr. 

  \item As clusters age, they spend a smaller fraction of time in 
  regions of high tidal strength. For example, in regions with $\lambdam > 3\times10^4$~Gyr$^{-2}$, this fraction decreases from $\sim$20\% immediately after formation to less than 1\% after 1~Gyr (Figure~\ref{fracinwriteout_aver}).
  
  \item {\it Ex situ} star clusters are brought in to the main galaxy by accretion of satellite galaxies. They experience similar tidal strength to the {\it in situ} clusters in the first $100$~Myr after formation, independent of their host galaxy mass (Figure~\ref{lambda_m_Mhalo}). After accretion {\it ex situ} clusters experience typically weaker tidal field than the {\it in situ} ones (Figure~\ref{sum_P_Mh}). They also tend to end up in higher energy orbits than the {\it in situ} clusters (Figure~\ref{RiRf}).

  \item Most {\it in situ} clusters remain near the galaxy center, but some are ejected to radii as large as 20 kpc on highly eccentric orbits, where they experience weaker tidal field (Figure~\ref{orb_eg}).

  \item The cluster bound fraction at the last available snapshot increases with initial cluster mass, meaning that lower mass clusters are more easily disrupted. At the same initial mass, the final bound fraction is usually higher for {\it ex situ} clusters than {\it in situ} clusters (Figure~\ref{sum_fb_Mi}).
  
  \item Clusters formed in more massive halos on average experience stronger tidal field. The difference is larger for {\it ex situ} clusters. For these clusters, the average normalization of the disruption rate 
  %$\bar{P}$ (Eqs.~\ref{eq:ttid}--\ref{eq:p}) 
  increases by a factor of $\sim$3 as host halo mass increases from $3\times 10^9$ to $10^{11}\Msun$ (Figure~\ref{sum_P_Mh}).

  \item \revv{We also use an alternative definition for the tidal strength $\Omega_{\rm tid}^2=\lambda_{\rm 1,e}$ and compare with our default choice $\lambdam$. Clusters experience stronger tidal field and end up with smaller bound fraction using this alternative, but our conclusions remain qualitatively the same (Appendix~\ref{appendix2}).}
  \revvv{Note that both methods provide only rough estimates for the actual cluster mass loss and may be quantitatively inaccurate. Better understanding and parametrization of cluster disruption is urgently needed.}
 \end{itemize} 

These results are important for calibrating the disruption rates of globular cluster in semi-analytical modeling of their evolution. Even the highest-resolution current cosmological simulations are unable to capture the detailed dynamics affecting cluster disruption. Therefore, such models still require additional boost of the disruption rates. \citet{cyt} present an example of such a calibration that allows it to match observed properties of globular cluster systems.

\section*{Acknowledgements}

 We thank Hui Li for producing the simulation suite used in this study, and for useful discussions. 
 We used python packages {\tt NumPy} \citep{harris2020array}, {\tt Matplotlib} \citep{Hunter:2007} and {\tt yt} \citep{yt}. 
 OG and XM were supported in part by the U.S. National Science Foundation through grant 1909063.

\section*{Data Availability}
 
 The data underlying this article will be shared on reasonable request to the corresponding author. 

\bibliographystyle{mnras}
\bibliography{gc,bzd,diskplus,tidal,ack}

\appendix 

\section{Bound fraction correction} \label{appendix}

 During the runtime of the \citet{li_gnedin19} simulations the disruption timescale $t_{\rm tid}^{\rm out}$ was calculated with an old normalization factor of $41.4\,{\rm Gyr}^{-1}/\Omega_{\rm tid}$ instead of $100\,{\rm Gyr}^{-1}/\Omega_{\rm tid}$, which gave a factor 2.5 faster disruption.
 Also the value of $\lambdam$ was overestimated by the expansion scale factor $a_{\rm box}(t)$. These two effects were corrected by \citet{li_gnedin19} in post-processing of the simulation outputs.
 We describe how we correct for these two factors in this appendix. 

 The disruption timescale used during the simulation differs from the more recent estimate (Equation~\ref{eq:ttid}) by a factor
 $$ 
    \frac{t_{\rm tid}^{\rm out}}{t_{\rm tid}^{\rm true}} 
    \equiv \omega = 0.414\,a_{\rm box}.
 $$
 The analysis presented in \citet{li_gnedin19} corrected for this overestimate of the disruption rate by recalculating the bound fraction in post-processing of the simulation outputs. They calculated the tidal tensor for each cluster at a given snapshot $n$ and applied the true timescale $t_{\rm tid}^{\rm true}(M_n,\Omega_n)$ to obtain a corrected value of the bound fraction at the next snapshot.

 Since we showed that the tidal field along cluster trajectories varies rapidly between the snapshots separated by a relatively large interval of $\sim 150$~Myr, we may expect the above correction to be inaccurate. The largest uncertainty comes from the timing of snapshots: the tidal field at a snapshot is at a random phase of the cluster orbit, and probably in a weaker tidal field region because clusters on eccentric orbits spend more time farther from the galaxy center.
 
 We can check the accuracy of this correction using the SFE200w run with very finely spaced output of the tidal field. That output matches the frequency of update of the bound fraction in the simulation runtime, once every global simulation timestep on the root grid level. If the number of timesteps between two snapshots is $m$, then we can write the calculation of $f_{\rm dyn}$ from one snapshot to the next as
 $$ 
    \frac{f_{n+m}}{f_{n}} = \exp\left(-\sum_{i=n+1}^{n+m} \frac{{\rm d}t_i}{t_{\rm tid}(M_i,\Omega_i)}\right).
 $$
 We corrected the tidal tensor history from the SFE200w run by the factor $\omega$ and calculated the true bound fraction of clusters in that run. Then we repeated the method used by \citet{li_gnedin19}, which also includes the factor $\omega$ but applies it over the whole duration between the snapshots at once, $\Delta t_n = \sum_{i} {\rm d}t_i$, instead of at $m$ individual steps. We found that using the coarser steps usually overestimates the bound fraction by $\sim$3\% for final $f_{\rm dyn} > 0.9$ and up to 10\% for final $f_{\rm dyn} \lesssim 0.8$. The more mass clusters lose, the more inaccurate this recalculation becomes.
 
 Here we derive an alternative correction for the bound fraction in the main runs, using directly the output values $f_{\rm dyn}^{\rm out}$ at the snapshots. The disruption timescale used in the runtime differs from the true one in two aspects: the normalization of the disruption rate was overestimated, and in turn it lead to underestimating current bound cluster mass $M_i$. In terms of the ratio of the output cluster mass to its true value, $\mu = f^{\rm out} / f^{\rm true}$, the disruption time used in the simulations can be written as 
 $$
    t_{\rm tid}^{\rm out} = t_{\rm tid}^{\rm true}(M\mu,\Omega/\omega) = \omega\, \mu^{2/3} t_{\rm tid}^{\rm true}(M,\Omega).
 $$ 
 Then the logarithmic difference of the bound fractions in the outputs is
 \begin{eqnarray}
    \log\left(\frac{f_{n+m}}{f_{n}}\right)^{\rm out}
    = -\sum_{i=n+1}^{n+m} \frac{{\rm d}t_i}{\omega_i\, \mu_i^{2/3} t_{\rm tid}(M_i,\Omega_i)} \nonumber\\
    \approx - \frac{1}{\bar\omega\, \mu_n^{2/3}} 
    \sum_{i=n+1}^{n+m} \frac{{\rm d}t_i}{t_{\rm tid}(M_i,\Omega_i)}
    = \frac{1}{\bar\omega\, \mu_n^{2/3}} \log\left(\frac{f_{n+m}}{f_{n}}\right)^{\rm true} \nonumber
\end{eqnarray}
 where $\bar\omega$ is taken as a simple average between the two snapshots $\bar\omega = 0.414 (a_{{\rm box},n} + a_{{\rm box},n+m})/2$, and $\mu_n = f_n^{\rm out} / f_n^{\rm true}$ is calculated from the quantities already known at snapshot $n$. Therefore, we correct the output fractions as
 $$
    \log\left(\frac{f_{n+m}}{f_{n}}\right)^{\rm true} = \bar\omega\, \mu_n^{2/3} \log\left(\frac{f_{n+m}}{f_{n}}\right)^{\rm out}.
 $$
 For the first snapshot where a new cluster appears, the corrected and uncorrected bound fractions $f_n^{\rm true}$ and $f_n^{\rm out}$ are both taken to be 1.
  
 Essentially this new method assumes an average $a_{\rm box}$ and an average mass correction factor. Although still an approximation, it encodes the evolution of the tidal field between snapshots via the available output values of the bound fraction. We checked this method on the SFE200w run with high time resolution outputs and found very good agreement within 1\%.

\section{An alternative definition of tidal strength} \label{appendix2}

 As an alternative, we introduce another parametrization of the tidal strength 
 $\lambda_{\rm 1,e}=\lambda_1-0.5(\lambda_2+\lambda_3)$ to describe 
 the tidal strength. 
 Here $\lambda_1, \lambda_2$ and $\lambda_3$ are the three eigenvalues 
 of the tidal tensor from the largest to the smallest. 
 The tidal field strength that sets the tidal radius of a cluster 
 on a circular orbit is $\partial^2\Phi/\partial r^2 + \Omega^2$ \citep{Renaud:2011aa}. 
 \citet{pfeffer_etal18} show in their Appendix~C that for a Plummer sphere 
 the tidal strength is well approximated by $\lambda_1-0.5(\lambda_2+\lambda_3)$. 
 Thus, we take this $\lambda_{\rm 1,e}$ as our alternative tidal strength, 
 as is done in \citet{Rodriguez:2022aa}. 
 We find the value of $\lambda_{\rm 1,e}$ to be always positive 
 and within a factor of 2 of the value of $\lambdam$. 
 Using this new $\lambda_{\rm 1,e}$, we take 
 $\Omega_{\rm tid}^2=\lambda_{\rm 1,e}$ instead of the original 
 $\Omega_{\rm tid}^2=\lambdam/3$ in the disruption calculation. 

\begin{figure}
  \centering
  \includegraphics[width=\columnwidth]{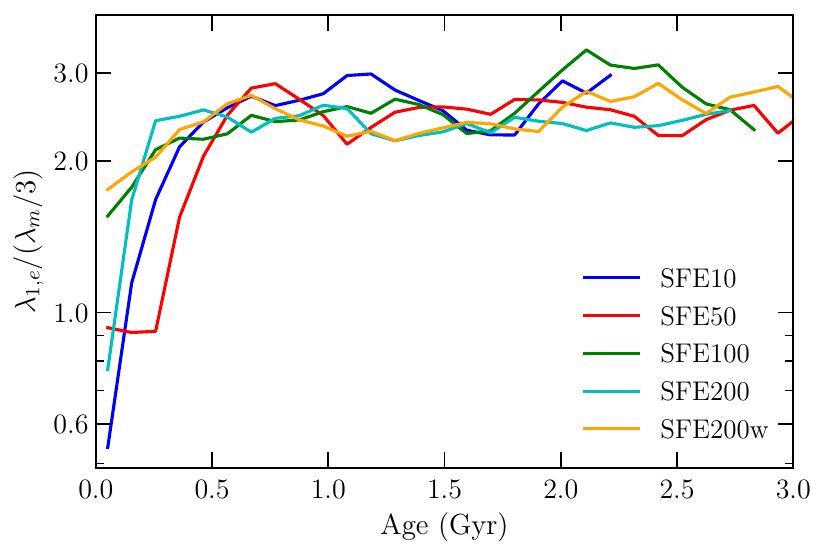}
  \caption{Comparison of $\lambda_{\rm 1,e}$ with $\lambdam/3$. 
           The y axis is the ratio of the median of the two $\lambda$ in age bins, 
           i.e. ratios of the tidal strength evolution as in the bottom left panel of Figure~\ref{sum_medevol}. 
           The median evolution of $\lambda_{\rm 1,e}$ is about 2-3 times larger than $\lambdam/3$. 
           Note that although the ratio between $\lambda_{\rm 1,e}$ and $\lambdam/3$ 
           is smaller when clusters are young, $\lambda_{\rm 1,e}$ is still larger 
           at the beginning than it is at later times. } \label{comp_medevol}
 \end{figure}

\begin{figure}
  \centering
  \includegraphics[width=\columnwidth]{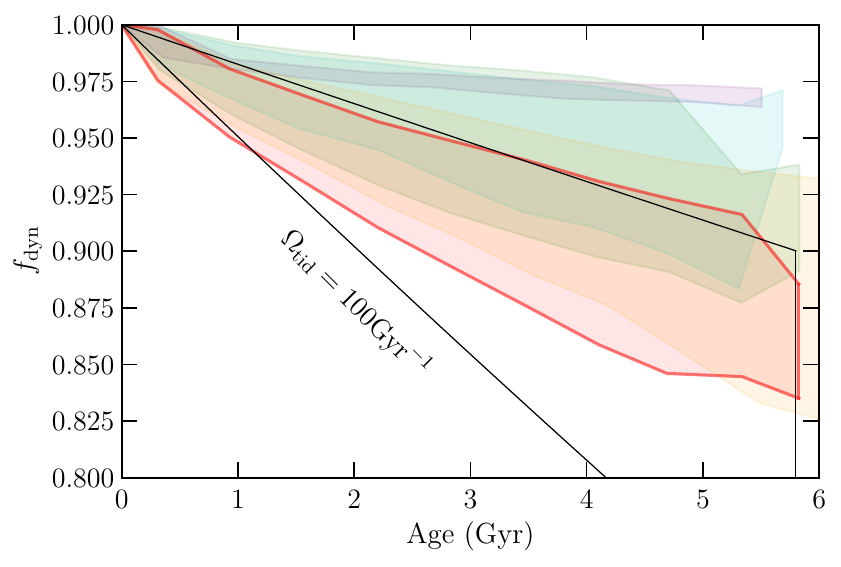}
  \caption{A reproduction of Figure~\ref{tidal_writeout_main_fdyn_age}, 
           but with $\Omega_{\rm tid}^2=\lambda_{\rm 1,e}$. 
           The bound fraction goes down with this alternative, 
           but the distinction for clusters from different origins remains similar. } \label{shade_1e}
 \end{figure}

 We show a comparison between the median evolution with cluster age 
 of the two in Figure~\ref{comp_medevol}. 
 In general, the median evolution of $\lambda_{\rm 1,e}$ is about 
 2-3 times larger than $\lambdam/3$. 
 The evolutionary trend of $\lambda_{\rm 1,e}$ is similar to that of $\lambdam$: 
 it has higher value at young ages and then plateaus to a lower value at late times. 
 The time fraction that clusters experience high $\lambda_{\rm 1,e}$ 
 also goes down with cluster age. 
 Since with this alternative definition of tidal strength 
 clusters experience stronger disruption, they end up with 
 smaller bound fraction in the end. 
 We show the evolution of bound fraction for clusters from different origins 
 in the SFE200w run, similar to Figure~\ref{tidal_writeout_main_fdyn_age}, 
 but calculated with $\Omega_{\rm tid}^2=\lambda_{\rm 1,e}$ 
 instead of the original $\Omega_{\rm tid}^2=\lambdam/3$ in Figure~\ref{shade_1e}. 
 Our conclusion that the bound fraction for clusters from satellites 
 usually decreases more slowly than for clusters formed in the main halo, 
 remains unchanged with this alternative.
 
\bsp	% typesetting comment
\label{lastpage}
\end{document}